\documentclass[a4paper,12pt]{article}
\pdfoutput=1
\usepackage{graphicx,rotating}
\usepackage{hyperref}\usepackage{slashed}
\def\hhref#1{\href{http://arxiv.org/abs/#1}{#1}} % in bibliography

\hypersetup{colorlinks,bookmarksopen,bookmarksnumbered,
linkcolor=blus,pdfstartview=FitH,urlcolor=rossos,citecolor=verdes}

\newcommand{\fig}[1]{~\ref{fig:#1}}
\usepackage{multicol}
\usepackage{color}
\definecolor{rosso}{cmyk}{0,1,1,0.4}
\definecolor{rossos}{cmyk}{0,1,1,0.55}
\definecolor{rossoc}{cmyk}{0,1,1,0.2}
\definecolor{blu}{cmyk}{1,1,0,0.3}
\definecolor{blus}{cmyk}{1,1,0,0.6}
\definecolor{bluc}{cmyk}{1,1,0,0.1}
\definecolor{verde}{cmyk}{0.92,0,0.59,0.25}
\definecolor{verdec}{cmyk}{0.92,0,0.59,0.15}
\definecolor{verdes}{cmyk}{0.92,0,0.59,0.4}
\newcommand{\nubarnu}{\raisebox{1ex}{\hbox{\tiny(}}\overline\nu\raisebox{1ex}{\hbox{\tiny)}}\hspace{-0.5ex}}

\newcommand{\riga}[1]{\noalign{\hbox{\parbox{\textwidth}{#1}}}\nonumber}

\newcommand{\eq}[1]{~{\rm (\ref{eq:#1})}}

\newcommand{\MeV}{\,{\rm MeV}}
\newcommand{\GeV}{\,{\rm GeV}}
\newcommand{\TeV}{\,{\rm TeV}}
\newcommand{\cm}{\,{\rm cm}}

\def\circa#1{\,\raise.3ex\hbox{$#1$\kern-.75em\lower1ex\hbox{$\sim$}}\,}

\newcommand{\lambdaN}{\lambda}

\newcommand{\beq}{\begin{equation}}
\newcommand{\eeq}{\end{equation}}

\font\tenrsfs=rsfs10 at 12pt
\font\sevenrsfs=rsfs7
\font\fiversfs=rsfs5
\newfam\rsfsfam
\textfont\rsfsfam=\tenrsfs
\scriptfont\rsfsfam=\sevenrsfs
\scriptscriptfont\rsfsfam=\fiversfs
\def\mathscr#1{{\fam\rsfsfam\relax#1}}

\def\Lag{\mathscr{L}}

\def\circa#1{\,\raise.3ex\hbox{$#1$\kern-.75em\lower1ex\hbox{$\sim$}}\,}
\makeatletter

\def\art{\@ifnextchar[{\eart}{\oart}}
\def\eart[#1]#2#3#4#5#6{{\rm #2}, {#3 #4} {\rm (#6) #5} [arXiv:{\hhref{#1}}]}
\def\hepart[#1]#2{{\rm #2, arXiv:\hhref{#1}}}
\newcounter{alphaequation}[equation]
\def\thealphaequation{\theequation\hbox to
0.6em{\hfil\alph{alphaequation}\hfil}}
\def\eqnsystem#1{
\def\@eqnnum{{\rm (\thealphaequation)}}
\def\@@eqncr{\let\@tempa\relax \ifcase\@eqcnt \def\@tempa{& & &} \or
 \def\@tempa{& &}\or \def\@tempa{&}\fi\@tempa
 \if@eqnsw\@eqnnum\refstepcounter{alphaequation}\fi
\global\@eqnswtrue\global\@eqcnt=0\cr}
\refstepcounter{equation} \let\@currentlabel\theequation \def\@tempb{#1}
\ifx\@tempb\empty\else\label{#1}\fi
\refstepcounter{alphaequation}
\let\@currentlabel\thealphaequation
\global\@eqnswtrue\global\@eqcnt=0 \tabskip\@centering\let\\=\@eqncr
$$\halign to \displaywidth\bgroup \@eqnsel\hskip\@centering
$\displaystyle\tabskip\z@{##}$&\global\@eqcnt\@ne
\hskip2\arraycolsep\hfil${##}$\hfil& \global\@eqcnt\tw@\hskip2\arraycolsep
$\displaystyle\tabskip\z@{##}$\hfil
\tabskip\@centering&\llap{##}\tabskip\z@\cr}
\def\endeqnsystem{\@@eqncr\egroup$$\global\@ignoretrue} \makeatother

\newcommand{\eV}{\,{\rm eV}}

\newcommand{\SU}{\,{\rm SU}}
\begin{document}

\begin{center}
IFUP-TH/2008-9\bigskip\bigskip\bigskip\bigskip

{\Huge\bf\color{red}Type-III see-saw at LHC}\\

\medskip
\bigskip\color{black}\vspace{0.6cm}
{
{\large\bf Roberto Franceschini}$^a$,
{\large\bf Thomas Hambye}$^b$,
{\large\bf Alessandro Strumia}$^c$
}
\\[7mm]
{\it $^a$ Scuola Normale Superiore and INFN, Pisa, Italy}\\[3mm]
{\it $^b$ Service de Physique Th\'eorique, Universit\'e Libre de Bruxelles, Belgium}\\[3mm]
{\it $^c$ Dipartimento di Fisica dell'Universit{\`a} di Pisa and INFN, Italia}

\bigskip\bigskip\bigskip

{\large
\centerline{\large\bf Abstract}

\begin{quote}
Neutrino masses can be generated by fermion triplets with TeV-scale mass,
that would  manifest at LHC as production of  two leptons together with two heavy SM vectors or higgs,
giving rise to final states such as $2\ell+4j$ (that can violate lepton number and/or lepton flavor)
or $\ell +4j +\slashed{E}_T$.
We devise cuts to suppress the SM backgrounds to these signatures.
Furthermore, for most of the mass range suggested by neutrino data, triplet decays are detectably displaced from
the production point, allowing to infer the neutrino mass parameters.
We compare with LHC signals of type-I and type-II see-saw. 
\end{quote}}

\end{center}

\newpage

\section{Introduction}

%The possibility that the mechanism of neutrino mass generation can be tested at LHC.
The smallness of the observed neutrino masses~\cite{review} suggests that they are more likely
generated at  an energy scale many orders
of magnitude above the TeV scale that is going to be explored by the LHC $pp$ collider.
Nevertheless  nothing prevents that the particles that mediate neutrino masses have TeV scale masses,
and it is worth to  see what would be their manifestations, given that 
finding  a signal of new physics among the backgrounds present at LHC 
is easier if one knows what to search for.
Tree level exchange of 3 different types of new particles with masses $M$ can generate neutrino masses:
\begin{itemize}

\item Type I see-saw employs at least two neutral fermions, the `right-handed neutrinos'
with Yukawa couplings $\lambda$:
in view of $m_\nu = \lambda^2 v^2/M$,
for $M\sim \TeV$ the Yukawa coupling $\lambda$ directly related to neutrino masses
are so small that right-handed neutrinos
are negligibly produced at LHC~\cite{typeI}.

\item Type II see-saw employs a $\SU(2)_L$ scalar triplet,
that would be produced at LHC via its gauge interactions 
leading to the signals explored in~\cite{LHCtypeIIa,LHCtypeII} if $M\circa{<}\TeV$;
however such a small $M$ is not technically natural. 
%and would open another hierarchy problem.

\item Type III see-saw \cite{foot} employs at least two $\SU(2)_L$ fermion triplets, and we here explore their LHC phenomenology.
A TeV-scale $M$ is technically natural, although unmotivated.

\end{itemize}
Combinations of these neutrino mass sources are also possible, as for example in left-right models which contains both type-I and type-II, or as with the adjoint representation of SU(5) which contains both type-I and type-III \cite{ma1,bajc1}.

Section 2 describes the triplet fermion Lagrangian.
Section 3 studies triplet production at LHC.
Section 4 studies triplet decays.
Section 5 combines the previous results to derive signatures at LHC.
In section 6 we present our conclusions.

%\footnote{
%[Type III see-saw employs a fermion triplet, so it is similar to the wino
%and to the $n=3$, $Y=0$ Minimal Dark Matter, plus an extra small coupling
%that completely changes the phenomenology at LHC.
%Leptogenesis should also be reconsidered if one considers the region $M \sim v$.]}

%%%%%%%%%%%%%%%%%%%%%%%%%%%%%%%%%%%%%%%%%%%%%%%%
\section{The Lagrangian}
Generic neutrino masses can be mediated by at least three fermion SU(2)$_L$ triplets $N^a$ with zero hypercharge:
the Lagrangian keeps the same structure as in the singlet case,
but with different contractions of the SU$(2)_L$ indices that we  
explicitly show:
\beq\label{eq:Lseesaw3}
\Lag = \Lag_{\rm SM} +\bar N_i iD\hspace{-1.5ex}/\, N_i +
\bigg[\lambdaN ^{ij} ~ N^a_i  
(L_j\cdot \varepsilon \cdot \tau^a \cdot H)  + \frac{M_{ij}}{2}   N_i^a N_j^a+\hbox{h.c.}\bigg].  \eeq
The $\SU(2)_L$ gauge index $a$ runs over $\{1,2,3\}$,
$\tau^{a}$ are the Pauli matrices
and $\varepsilon$ is the permutation tensor
($\varepsilon_{12}=+1$);
$i,j$ are flavor indices.
Gauge covariant derivatives are defined as $D = \partial + i g A^A T^A$;
the lepton doublets $L= (\nu,e_L)$ and the Higgs doublet $H= (h_+,h_0)$ 
have hypercharge $Y=-1/2$ and $Y=1/2$ respectively and transform as a 2 under $\SU(2)_L$.
Taking into account the breaking of $\SU(2)_L$, we choose the usual unitary
gauge where $h_+=0$
and $h_0 = v +h/\sqrt{2}$ with $v=174\GeV$.
The three components of the Weyl fermion
$N^a$ are $N^0 \equiv N^3$ with charge zero and
$N^\pm \equiv (N^1\mp i N^2)/\sqrt{2}$ with charge $\pm 1$.
With this convention all kinetic and mass terms have conventional normalization.
%\begin{eqnarray}
%N^a N^a &=& N_0^2 + 2 N_+ N_-\\
%\bar N i \slashed{D} N &=&  \nonumber
%(\bar N _0 i \ds N _0 + \bar N _+ i \ds N _+ + \bar N _- i\ds N _-)\\
%&&-
%g_2 s_W (\bar N _+ \slashed{A} N _+ - \bar N _- \slashed{A} N _-) -
%g_2 c_W (\bar N _+ \slashed{Z} N _+ - \bar N _- \slashed{Z} N _-)+ \\
%&&
%-g_2(\bar N _- \slashed{W}^- N _0 -\bar N _0 \slashed{W}^- N _+)
%+ g_2 (\bar N _+ \slashed{W}^+ N _0 - \bar N _0\slashed{W}^+ N _-)\nonumber\\
%N^a  
%(L \cdot \varepsilon\cdot \tau^a \cdot H)&=&-
%N_0 (e_L h_+ + \nu h_0) + \sqrt{2}  N_- \nu h_+ - \sqrt{2}
%N_+ e_L h_0
%\end{eqnarray}
Defining a Dirac spinor $\Psi_C= (N_-, \bar N_+)$ 
%(such that the particle has charge $-e$ and the anti-particle has charge $+e$)
and a Majorana spinor $\Psi_N=(N_0,\bar N_0)$ the vector couplings become all of vectorial type \cite{ABBGH2}:
\beq\label{eq:Lag3}\Lag=
-g_2 (\bar\Psi_N \slashed{W}^+\Psi_C + \bar\Psi_C \slashed{W}^-\Psi_N)+e \bar\Psi_C \slashed{A} \Psi_C +{g_2}{c_W} \bar\Psi_C \slashed{Z} \Psi_C+\cdots \eeq
$N_0$ has the same Yukawa interaction and mass term as the right-handed neutrino of type I see-saw:
so one has the usual see-saw formula for 
light Majorana neutrino masses 
${m}_\nu = - (v{\lambda} )^T  \cdot {M}^{-1}\cdot (v{\lambda})$.
We work in the basis of $N_i$ mass eigenstates, where $M_{ij} = {\rm diag}\,(M_1,M_2,M_3)$
with $0<M_1\le M_2 \le M_3$.
The light neutrino masses are $0\le m_1 \le m_2\le m_3$.
We define the parameters
$\tilde{m}_i \equiv |\sum_j \lambda_{ij}^2 v^2/M_i|$ that 
tell how fast $N_i$ decays: $\Gamma_i=\tilde{m}_i M_i^2 /(8 \pi v^2)$.
%The flavor structure of $\lambda_{ij}$ as well as the 
The $\tilde{m}_i$ are unknown, but must satisfy the following neutrino mass constraints
\beq
\tilde{m}_i \geq m_1,\qquad
\sum_i\tilde{m}_i \geq \sum_{i=1}^3 m_{i}  . \label{eq:boundsum}
\eeq
The first bound is known \cite{BP1} and only holds with three triplets.
The second bound can be derived applying the Schwarz inequality to the trace of  the neutrino mass matrix.
%using the parameterization of the Yukawa couplings $\lambda_{ij}$
The parameter $\tilde{m}_1$ is unknown:  it can be comparable to the
observed solar and atmospheric mass splitting, or it could be much smaller (or much larger if cancellations between large Yukawa couplings occur in the neutrino masses).

One or more $N_i$ could be light enough to be probed by LHC.

It would be especially interesting to measure the properties of the lightest triplet, $N_1$,
that could play an important r\^ole in cosmology: its decays source a lepton asymmetry, dominantly
at temperatures $T\sim M_1/20$,
somewhat depending on the precise interplay between the annihilation rate, the decay rate and the expansion rate~\cite{LeptogenesisTypeIII}.
According to the SM, in cosmology a non-zero Higgs vev starts to 
appear via a second-order phase transition below
a critical temperature comparable to the Higgs mass $m_h$, suppressing the sphaleron rate~\cite{shap}.
As a consequence thermal
baryogenesis via leptogenesis is suppressed if $M_1\circa{<}10 m_h$,
leading to some absolute boundary on $M_1$ even in the most favorable case of a large CP asymmetry
in $N_1$ decays. Along the lines of~\cite{Hall} one can invent arguments that allow to argue that
values of $M_1$ around this anthropic boundary  are motivated by multiverse considerations.

\medskip

We focus on the lightest heavy triplet $N_1$, 
we denote as $\ell$ the unknown combination of $e,\mu,\tau$ coupled to $N_1$,
and from now on we drop the index $_1$ on $M$ and $N$.
%and we define its contribution to
%neutrino masses as $\tilde m_1 \equiv (\lambda_1 v)^2/M$.
The coupling to the Higgs generates a mass-mixing term between $N_0, N_\pm$ and $\nu_\ell ,\ell_L$ respectively:
%Type III see saw involves the charged $N^\pm$ with its electroweak gauge couplings and the
%coupling to the Higgs:
%\beq \Lag \supset -\sqrt{2}\lambdaN v N_+ e_L  -\lambdaN h N_+ e_L + \hbox{h.c.}\eeq
%The first mass term, together with the charged lepton mass term gives a mass mixing 
%between SM leptons and $N_\pm$:
\beq
\bordermatrix{& \nu_\ell & N_0 \cr \nu_\ell & 0 & -{\lambda} v \cr N_0 & -{\lambda} v & M},\qquad
\bordermatrix{&\ell_L & N_- \cr \ell_R& {\lambda}_\ell v &0 \cr N_+ &- \sqrt{2}{\lambda} v & M}\eeq
%where bold-face reminds that $\mb{\lambda}_N$ and $\mb{M}_N$ are $3\times 3$ flavour matrices.
The resulting $N_0/\nu_\ell$ and $N_-/\ell_L$ mixings are
of order  $\lambda v/M = \sqrt{\tilde{m}_1/M} \sim 10^{-6}$
and can be neglected
(the $Z$ couplings of $\ell_L$ has been measured at LEP with precision of about $10^{-3}$)
and the $N_0,N_\pm$ couplings to the Higgs $h$ remain unchanged.
The resulting mass splitting between the charged and the neutral components of the $N^a$ multiplet 
is of order $\tilde{m}_1$ and can be neglected
%The correction to the $N_\pm/N_0$ mass splitting is also negligible:
%\beq M_+ \simeq  M +\frac{\tilde{m}_1}{2} + \cdots\eeq
with respect to the mass splitting generated by one-loop corrections:
in the limit $M \gg M_Z$ one has $\Delta M\equiv M_\pm - M_0= 166\MeV$~\cite{MDM}.

The only relevant effect of the mixings is to generate $\SU(2)_L$-breaking couplings between
the $N^a$, the SM leptons and the heavy $Z,W^\pm$ vectors that `eat' the Goldstone components of the Higgs.
This is how the $\lambda~NLH$ couplings involving the Goldstones reappear in the Lagrangian.
Indeed, the unitary field redefinitions that define the mass eigenstates at first order in $\lambdaN v/M\ll1 $
\beq \nu_\ell \to \nu_\ell -\frac{\lambdaN v}{M} N_0,\qquad
N_0\to N_0 +\frac{\lambdaN v}{M} \nu_\ell,\qquad 
\ell_L \to \ell_L - \sqrt{2}\frac{\lambdaN v}{M} N_-,\qquad
N_-\to N_- +\sqrt{2}\frac{\lambdaN v}{M} \ell_L \label{eq:mixing}\eeq
generate the following couplings~\cite{Bajc,ABBGH2}
%give rise to small SU(2)-breaking couplings of triplets to vectors that allow triplet decays:
\beq \Lag\supset \frac{g_2}{2c_W}\frac{\lambda v}{M}  Z_\mu (\bar N_0 \gamma_\mu \nu + \sqrt{2} \bar N_- \gamma_\mu e_L + \hbox{h.c.}) + \frac{g_2}{2}\frac{\lambda v}{M}  W^+_\mu (2 \bar N_+ \gamma_\mu \nu - \sqrt{2}\bar N_0 \gamma_\mu e_L)+\hbox{h.c.} \eeq
%No couplings to $A_\mu$ are generated, because electromagnetism is unbroken.
relevant for $N_0,N_\pm$ decays. Such interactions are typical of models where heavy states
have a SU(2)$_L$-breaking mixing with leptons. 
Type-III interactions induce mixing effects not only for the (unobservable) neutrinos like type-I models but also for charged leptons which leads to a rich phenomenology.

Precision electroweak data are affected via a small correction to the $W$ parameter of~\cite{STWY}:
$W= +\alpha_2 M_W^2/15\pi M^2$.

\begin{figure}[t]
\begin{center}
\includegraphics[width=0.45\textwidth]{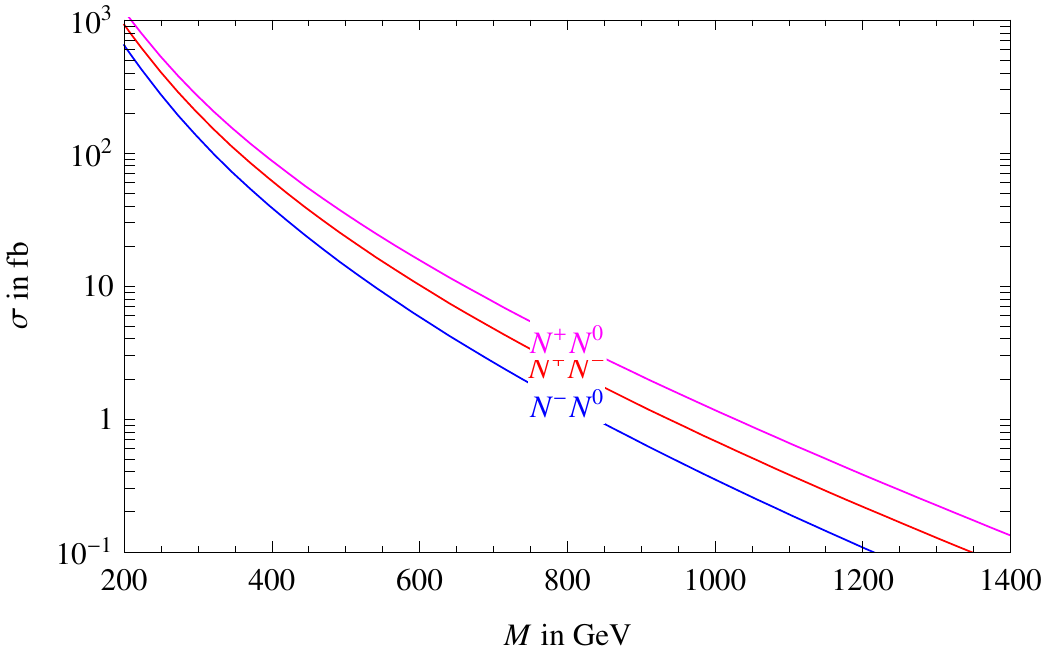}\quad
\includegraphics[width=0.45\textwidth]{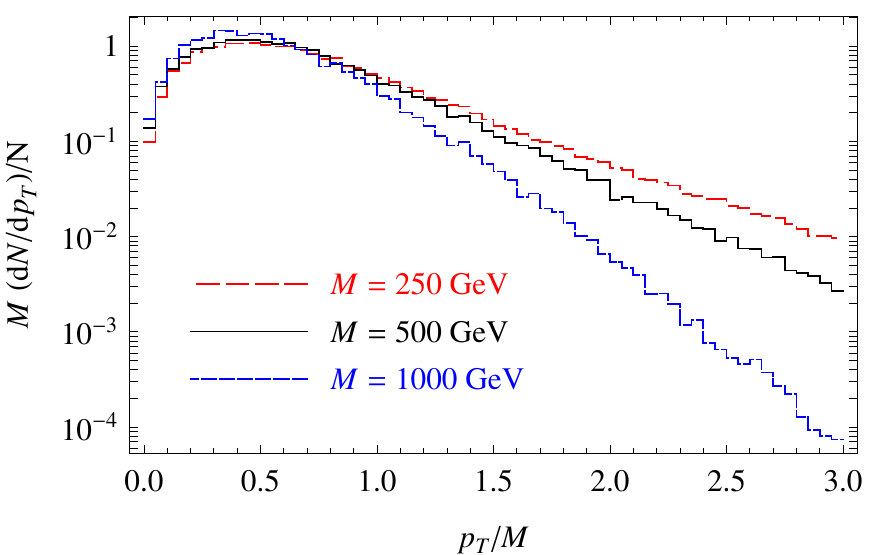}
\caption{\label{fig:sigma}\em Left plot: total production cross section at LHC.
Right plot: $p_T$ distribution.}
\end{center}
\end{figure}

%%%%%%%%%%%%%%%%%%%%%%%%%%%%%%%%%%%%%%%%%%%%%
\section{Production at LHC}
Production at LHC is dominated by gauge couplings:
the dominant partonic process that leads to pair production of triplets in $pp$ collisions is
$q\bar q' \to W^\pm, Z$.
%This was already computed in the limit $\lambda=0$ in~\cite{MDM1}.
%The gauge couplings to the $W^+$, the $W^-$ and the $Z$ of the neutral component of an
% $n$-plet of SU(2) with hypercharge $Y$ are:
%$$ g_{\rm CC}^\pm=\frac{g_2}{\sqrt{2}} \cdot \frac{\sqrt{n^2-(1\mp 2Y)^2}}{2},\qquad
%g_{\rm NC}=\frac{g_2 Y}{c_{\rm W}}.$$
%For $n=3$ and $Y=0$ one has $ g_{\rm CC}^\pm=g_2$ and $g_{\rm NC}=0$.
%The $Z$ coupling of the charged components is $g_2 c_W T_3 = g_2 c_W$.
%
%For $n=2$ and $Y=1/2$ one has the usual $ g_{\rm CC}^\pm=g_2/\sqrt{2}$ coupling of doublets.
The partonic production cross sections, summed over initial state colors and over final state polarizations,
and averaged over initial state polarizations, are
\begin{eqnsystem}{sys:sigmas} \frac{d\hat\sigma}{d\hat t}&=&\frac{V_L^2 + V_R^2} {16\pi \hat s^2} N_c (2M^4 + \hat s^2 - 4 M^2 \hat t + 2\hat s\hat t + 2\hat t^2),\\
\hat\sigma &=&\frac{\beta(3-\beta^2)}{48\pi }\label{eq:sigmatot}
N_c \hat s (V_L^2 + V_R^2)\end{eqnsystem}
where $N_c = 3$, $\beta\equiv \sqrt{1-4M^2/\hat s}$ is the $N$ velocity ($0\le\beta\le 1$) and
\beq\label{eq:V}\begin{array}{ll}
V_A=0 & \hbox{for $q\bar q\to N^0 N^0$ }\\
V_A= \displaystyle\frac{Q_q e^2}{\hat s}+
\frac{g_A^q g_2^2}{\hat{s} - M_Z^2}
\qquad& \hbox{for $q\bar q\to N^+ N^-$ }\\
V_A = \displaystyle\frac{g_2^2}{\hat s-M_W^2}\frac{\delta_{AL}}{\sqrt{2}}& \hbox{for $u\bar d\to N^+ N^0$ }
\end{array}\eeq
where $g_A^q = T_3 - s_W^2 Q_q$ is the $Z$ coupling of quark $q$ for $A=\{L,R\}$.
%The $W^\pm$ line also take into account the gauge couplings times a $\sqrt{2}$
%combinatorial 2 factor
%that differentiates $ZZ$ and $\gamma\gamma$ from $W^+W^-$.
This result does not agree with eq.~(10) of~\cite{Ma}.
$\SU(2)_L$ invariance is restored in the limit $M^2\gg M_Z^2$, and the result 
$2\hat\sigma_{u\bar u} =2 \hat \sigma_{d\bar d}=\hat \sigma_{u\bar d}=
\hat\sigma_{d\bar u}$
agrees with~\cite{MDM}.
The cross section  $e^-e^+\to N^+ N^-$, relevant for a possible future collider, is found 
by replacing $q\to e$ in eq.\eq{V}.

Fig.\fig{sigma}a shows $\sigma(pp\to N_0 N_\pm )$ and $\sigma(pp\to N_+ N_- )$ 
as function of $M$ at LHC, i.e.\ at $\sqrt{s}=14\TeV$.
We integrated the parton distribution functions of~\cite{pdf}, and we checked that the result numerically
agrees with the one obtained implementing the triplet model in {\sc MadGraph}~\cite{MADGRAPH}.
This would lead to about $3\cdot10^3$ (10) pairs created at LHC for $M=250$~GeV ($M=1$~TeV) for an integrated luminosity of 3/fb
which should be collected at LHC in less than one year. These numbers have to be multiplied by about 2 orders of magnitude after 5 years of data taking.
Therefore LHC should be able to produce at least a few tens of events up to masses of $M\sim 1\TeV$, or even $1.5$ TeV in the long term.
Fig.\fig{sigma}b shows the distribution in the transverse momentum $p_T$, as computed by our MonteCarlo
for three representative values of $M$: it is peaked at $p_T \sim M/2$.

%For later convenience the dashed line show, as function of the Higgs mass, the main cross section
%for Higgs production in $pp$ collisions at LHC.
%CE LA FAI A FARLO RAPIDAMENTE CON MADGRAPH?

%
%It does not agree with Ma, who would got $3+\beta^2$
%because he missed the interference. It agrees with analogous computations
%in the book by Barger, pag 120]

Testing the production cross section would allow to identify the quantum numbers of the particle
and to test that/if the theory correctly predicts
the gauge interactions of a fermionic $\SU(2)_L$ triplet
(e.g.\ for a scalar replace $\beta(3-\beta^2)\to \beta^3/2$ in eq.\eq{sigmatot})
but not to study its connection with neutrinos.
This is encoded in $N_0,N_\pm$ decays, that also define the signatures at LHC.

%%%%%%%%%%%%%%%%%%%%%%%%%%%%%%%%%%%%%%%%%%%%%%%%%
\section{Decays}
%About decays into vectors, we recall that the coupling $(g_2/\sqrt{2})(\bar  t\slashed{W}P_L  b)$
%gives (the $P_L$ contributes as $1/2$ in $\Gamma_t$)
%$$\Gamma_t = \frac{g_2^2 m_t}{64\pi} \frac{m_t^2}{M_W^2}(1-\frac{M_W^2}{m_t^2})^2(1+\frac{M_W^2}{m_t^2}) $$
The $N_0$ decay widths agree with the results of~\cite{Bajc}:
\begin{eqnsystem}{sys:N0decay}
\Gamma(N_0\to \nu_\ell h ) =\Gamma(N_0\to \bar\nu_\ell h ) &=& \frac{1}{8}\frac{\lambda^2 M}{8\pi}
(1- \frac{m_h^2}{M^2})^{2},\\
\Gamma(N_0\to Z \nu_\ell) =\Gamma(N_0\to Z\bar\nu_\ell) &=& \frac{1}{8}\frac{\lambda^2 M}{8\pi}
(1- \frac{M_Z^2}{M^2})^{2}(1+2\frac{M_Z^2}{M^2}),\\
\Gamma(N_0\to W^+ \ell^- ) =\Gamma(N_0\to W^- \ell^+ ) &=& \frac{1}{4}\frac{\lambda^2 M}{8\pi}
(1- \frac{M_W^2}{M^2})^{2}(1+2\frac{M_W^2}{M^2})
.\end{eqnsystem}
The Higgs contribution is only possible if $m_h < M$
and for $M\gg m_h$
it equals $1/8$ of the $\SU(2)_L$-invariant width, $\Gamma(N_0) =\Gamma(N_\pm)= \lambda^2 M/8\pi$.
The remaining terms arise because of lepton mixing, eq.~(\ref{eq:mixing}), and the fact that in the unitary gauge the $W^\pm,Z$ vectors became massive `eating'
completely the Goldstones in the Higgs doublet. 
% $h_+$ is the longitudinal component of the $W^+$, 
%and $\eta$ is the longitudinal component of the $Z$. 
%If instead $m_h > M$ the decays into $h$ are kinematically forbidden.
The possibility that $M < M_W$ such that all 2-body decays are kinematically forbidden
is already excluded by LEP2.

\begin{figure}[t]
\begin{center}
\includegraphics[width=\textwidth]{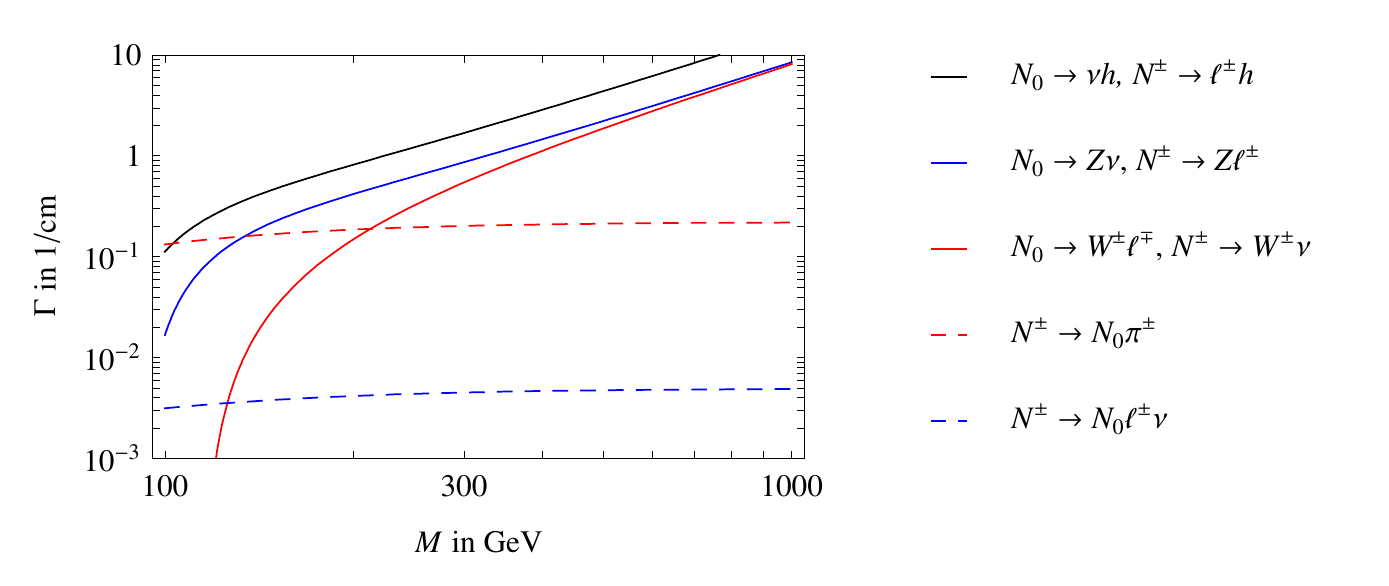}
\caption{\label{fig:Gamma}\em Triplet decay widths as function of the triplet mass for $\tilde{m}_1 = {\rm meV}$ and
$m_h=115\GeV$.}
\end{center}
\end{figure}

The $N_\pm$ decay widths are given by~\cite{Bajc}
\begin{eqnsystem}{sys:Npmdecay}
\Gamma(N^\pm \to \ell^\pm h ) &=& \frac{1}{4}\frac{\lambda^2 M}{8\pi}
(1- \frac{m_h^2}{M^2})^{2} , \\
\Gamma(N^\pm \to \ell^\pm Z ) &=& \frac{1}{4}\frac{\lambda^2 M}{8\pi}
(1- \frac{M_Z^2}{M^2})^{2}(1+2\frac{M_Z^2}{M^2})  , \\
\Gamma(N^\pm \to  \nubarnu_\ell W^\pm ) &=& \frac{1}{2}\frac{\lambda^2 M}{8\pi}
(1- \frac{M_W^2}{M^2})^{2}(1+2\frac{M_W^2}{M^2}).\\[3mm]
%.\end{eqnsystem}
%The new feature are $L$-violating decays due to $\lambda_N$: given that $\nubarnu$ cannot be detected,
%$L$-violation can be seen as $pp\to N^0N^0\to W^+W^+\ell\ell$
%and as $pp\to N^0 N^+X \to W^+ \ell \ell X$.
%HOW CAN ONE TAG THE $W^+\to \ell^+$?
%WHAT IS THE SM BACKGROUND?
\riga{The charged/neutral mass small splitting $\Delta M \approx 166\MeV$ is bigger than $m_\pi$ 
in all the allowed range of $M$ and gives rise to the following extra decay channels~\cite{MDM}}\\[3mm]
%\begin{eqnsystem}{sys:Npmdecay2}
\Gamma(N ^\pm \to N ^0 \pi^\pm ) &=& 
\displaystyle
\frac{2G_{\rm F}^2V_{ud}^2\, \Delta M^3 f_\pi^2}{\pi}
\sqrt{1-\frac{m_\pi^2}{\Delta M^2}},\\[3mm]
\Gamma(N ^\pm \to N ^0 e^\pm\nubarnu_e ) &=&\displaystyle
\frac{2G_{\rm F}^2 \,\Delta M^5}{15\pi^3} ,\\[3mm]
\Gamma(N ^\pm \to N ^0 \mu^\pm\nubarnu_\mu ) &=&0.12\ \Gamma(N ^\pm \to N ^0 e^\pm\nubarnu_e ) 
\end{eqnsystem}
which do not depend on any free parameter. 
Fig.\fig{Gamma} shows the various decay rates as function of $M$ for fixed $\tilde{m}_1 = {\rm meV}$;
we recall that $\lambda^2 M = \tilde{m}_1 M^2/v^2$.

Notice that, while $pp\to N^0 N^0$ does not arise at tree level,
this production channel is effectively produced by
the $N^\pm \rightarrow N^0 \pi^\pm$ decay 
because the $\pi^\pm$ are too soft to be observed.
The decay mode into pions is dominant for 
$\tilde{m} \circa{<}3\cdot 10^{-4}\,\hbox{eV}\cdot (100 \,\hbox{GeV}/M)^2$,
so that the effective production rate $pp \rightarrow N^0 N^0$ is given by the sum of all cross sections in fig.~1a.

%\left\{\begin{array}{ll}
%{G_{\rm F}^2 \Delta M^5}/{15\pi^3} &\hbox{if $X$ is a fermion}\\
%{2G_{\rm F}^2 \Delta M^5}/{135\pi^3} &\hbox{if $X$ is a scalar}
%\end{array}\right.
% having used the normalization $f_\pi = 131\,{\rm MeV}$~\cite{CHPT}.
%The computed  $\Delta M > m_\pi$
%(such that the two-body decay $N ^\pm \to N ^0 \pi^\pm $ is allowed)
%for any allowed value of $M \circa{>}100\GeV$.

%%%%%%%%%%%%%%%%%%%%%%%%%%%%%%%%%%%%%%%%%%%%%%%%
\section{Signals at LHC and displaced vertices}
Production of $N^0 N^\pm$ and $N^\pm N^\mp$
and their decays give rise to a variety of possible final states.
We focus on those involving jets, that have higher rates than purely leptonic final states,
and need a discussion of Standard Model backgrounds and how they can be suppressed.
In section~\ref{highest} we study the signal with the higher rate;
 lepton flavour violation (LFV) is studied in section~\ref{sec:LFV},
 and  lepton number violation (LNV) in section~\ref{LNV}.
%We prefer to discuss these discoveries using  channels suitable for the largest range of the parameter $M$.
%In what follows we leave implicit that events with each particle replaced by its anti-particle are also possible, besides we do not explicitly write the extra hadronic part of the interaction.
For simplicity in what follows we often leave implicit that events with each particle replaced by its anti-particle are also possible:
signal and background rates are similar but not equal.
%and we only list final state particles produced by the hard partonic $pp$ sub-processes.

%%%%%%%%%%%%%%%%%%%%%%%%%%%%%%%%%%%%%%%%%%%%%%%
\subsection{The signal with the higher rate}\label{highest}
In view of the small fb-scale cross sections for $N^0,N^\pm$ production, 
we first discuss the channel with the relatively higher rate
(for $\tilde{m}_1 \circa{>} 10^{-4}\eV$): 
\beq  pp\to N^+ N^0 \to \bar\nu W^+ W^\pm \ell^\mp
\to \hbox{4  jets + missing energy +
a charged lepton, all hard}. \label{signal}\eeq 
In the following discussions we will consider the following set of cuts, needed to
make the leptons and jets identifiable:
\beq
p_{T,j}>20 \textrm{GeV},~~~ p_{T,\ell}>10 \textrm{GeV},~~~ \Delta R_{\ell\ell,jj,j\ell}>0.4,~~~ \eta_{j,\ell}<5 . \label{gencuts}
\label{eq:gencuts}
\eeq
where $p_T$ is the momentum orthogonal to the beam axis,
$\eta$ is the pseudo-rapidity and $\Delta R=(\Delta\phi^2_T+\Delta\eta^2)^{1/2}$,
with $\Delta \phi_T$ being the angular separation in the plane $T$ransverse to the beam.
In order to suppress the SM backgrounds we will also implement the extra
cut of eq.~(\ref{mT}):  about 24\% (36\%)
of the signal events pass all cut requirements for $M=250~(750)$ GeV.
With an integrated luminosity of 3/fb,
this would lead to about $100$ events  for $M=250$~GeV.

\medskip

Disregarding mis-detections and particles escaping because of the geometrical acceptance of the detectors we identify as possible SM backgrounds: 
\begin{itemize}
\item[i)] $pp \to (V\to 2j)(V\to 2j)(W^-\to \ell \bar\nu)$ with $V$ either a $W$ or $Z$; its total cross section is about 37 fb~\cite{MADGRAPH}.

\item[ii)] $pp \to 4j(W^-\to \ell \bar\nu)$ where $j$ means a light jet orginated by QCD interactions
has a  larger cross section, about $4500$ fb~\cite{AlpGen}, after imposing the cuts of eq.\eq{gencuts}.

\item[iii)] $pp \to (t\to b(W^-\to\ell \bar\nu))(\bar{t}\to \bar bjj)$ has an even larger cross section, about 160 pb.

\end{itemize}
All above background cross sections have been computed summing over lepton flavors and
can be reduced to zero exploiting the fact that type-III see-saw produces
uncorrelated $\ell\bar{\nu}$ ($\bar{\ell}\nu$) pairs, while the above SM processes produce pairs correlated through $W$ decays. 
In practice, this can be done by  requiring that the measurable transverse mass $m_T$ of the $\ell,\bar\nu$ system
does exceed the $W$ mass, such that these particles cannot come from $W$ decays:
\beq 
	{m}_T^2\equiv2 E_T^\ell \slashed{E}_T(1-\cos\phi_{e\nu}^T)> M_W^2 \label{mT}
\eeq %il BR e' 0.25
where $\phi_{e\nu}^T$ is the angle between the missing transverse momentum  $\slashed{p}_T$
and the transverse component of the lepton momentum,
$p_T^\ell$.
Having used this cut,  it will be of utmost importance to understand the instrumental sources of  isolated leptons uncorrelated to the missing energy: we do not fully address this detector issue.
%We find that process like $4j(Z\to l^+l^-)$ with one lepton escaping detection are not a source of missing energy uncorrelated to the missing lepton, in fact the missing energy of this kind of events is the energy of the escaped lepton, which is correlated to the observed one through the mother particle, the Z. Therefore this kind of background can be entirely rejected putting the threshold of equation (\ref{mT}) beyond the Z mass. The rejection is potentially  harder when the particle generating the missing energy is not the same which generates the observed lepton. This happen for instance in $$4j W^\pm W^{\pm,\mp} \to 4j l^\pm l^{\pm,\mp}_{escaping}+\slashed{E}_T,$$ where $l_{escaping}$ is a not detected lepton and 4j means four QCD jets.
%This process cannot be simulated by any code known by us, however it's cross section has to be at worse of the same order the process with one less jet, that is $O(80fb)$ or less. If the lepton can be missed only because of geometrical acceptance this process should not threaten our conclusions. Indeed we simulated the same process with the four jets originated by the decays of weak vectors and we found that a lepton with $\eta>5$, that is a not detected lepton, is produced in a fraction of events smaller than $5\cdot 10^{-5}$. Assuming that the same fraction of leptons falls out of the coverage of the detector in the QCD process we reach the conclusion that this kind of BG has a limited impact of $O(10^{-3}$ fb) or less.
\begin{itemize}

\item[iv)] $pp\to 4j(Z\to \nu\bar\nu)(W^-\to\ell\bar\nu)$ has a cross section of about $200$ fb~\cite{Piccinini}
% 3.2pb *  20%  * 32%
(summed over lepton flavors; only the cuts on jets are imposed).
This background  cannot be suppressed using eq.~(\ref{mT}) and is not yet implemented
in any available code; therefore we can only study it in a semi-quantitative way.
We estimate that it can be reduced down to $\sim 10$ fb by requiring that the $4j$ system reconstructs a $WW$ pair,
having assumed that two jets $j_1$ ad $j_2$ reconstruct a vector $V$ if their invariant mass $m(j_1j_2)$ satisfies 
\beq |m(j_1j_2)-M_V|<10 \GeV\label{eq:jjW}.\eeq
Furthermore, 
taking into account that larger $M$ implies a smaller signal but also harder leptons and jets,
we can impose e.g.:
\beq 
{p_T^\ell}\circa{>}0.25 M. \label{hardlepton}
\eeq 
this cut on the lepton $p_T$ only mildly reduces the signal.

\end{itemize}

%Under the assumption of a perfect detector, the SU(2)$_L$ triplet $N$ is discoverable with just 
%a few events, that can be collected with an integrated luminosity $\mathcal{L}=0.1$/fb (300/fb)  if $M=250$ GeV ($1.4$ TeV). 
%simgatot(1.4TeV)=0.27fb, sigmaBR(1.4TeV)=0.0675 con i cut diventa 0.027fb che sono 2.7 eventi con 100/fb di lumi 

Observing the process in eq.~(\ref{signal}) allows to determine the parameter $M$.  Indeed $M$ is equal to the invariant mass of the system made out of the charged lepton and the two jets produced in the $N^0$ decay. To determine which two jets come from $N^0$ decay one can divide the four jets in pairs, each with invariant mass close to $M_W$, and assume that the right jets pair is the one with the smallest $\Delta R$ with respect to the charged lepton. 
%POTREI QUANTIFICARE QUAL'E' L'EFFICIENZA DI QUESTA PROCEDURA DI SELEZIONE, MA CI VUOLE FORSE UN GIORNO PER SCRIVERE IL CODICE.	
%As soon as mildly rarer channels appear, the most characteristic features are: lepton flavor violation, lepton number violation, and possibly displaced decay vertices.

After the discovery of the signal eq.~(\ref{signal}) other signatures with lower rate but similar final states can be used to check the model. Once the properties of the Higgs boson will be assessed by the LHC,  eqs. (\ref{sys:N0decay})  and  (\ref{sys:Npmdecay}) predict the ratio of any possible decay chain without  pions over the rate of the discovery signal. This  prediction is independent of the value of $\lambda$ and weakly sensitive to $M$ (for $M$ at least few hundreds of GeV), providing an early check on the coupling structure of the theory we are studying.

\bigskip

%%%%%%%%%%%%%%%%%%%%%%%%%%%%%%%%%%%%%%%%%%%%%%%
\subsection{Lepton-flavour violating signatures}\label{sec:LFV}
Lepton-flavour violation manifests in processes like  
$pp\to  \ell_1 \bar\ell_2 ZW^+$ and $pp\to \ell_1\bar \ell_2 Z Z,$ where $\ell_1$ and $\ell_2$ denote any two lepton flavors present in $\ell$,
whose branching ratios depend on the flavor of $\ell$. 
The decays $Z\to \mu^+\mu^-$ or $Z\to e^+ e^-$ would give the cleanest signature.
We here study the less clean signature with largest event rate, obtained when
all weak vectors decay hadronically, giving
\beq
pp \to \ell_1 \bar\ell_2 4j \label{LFV}
\eeq
without missing transverse energy $E_T$. %il BR di questo dec. e' 0.12
The SM backgrounds that can fake this signal are processes where
lepton flavor is carried away by neutrinos, that escape carrying away
a  missing $E_T$ below the energy resolution of the detector.
%
%%With a perfect detector this channel has no missing $E_T$, however this is happen to be unrealistic and a small amount $\delta$ of instruments-originated missing energy has to be taken into account. 
%Therefore  backgrounds that can fake the signal of eq.~(\ref{LFV})
%arise from SM processes where neutrinos are produced and escape
%carrying their lepton flavor and undetected 
%The presence of this instrumental source of missing energy is crucial to make a list of the possible standard model reactions that can fake the signal of Eq. (\ref{LFV}). 
One needs to consider at least the following processes
(we quote background cross-sections for any given pair of lepton flavors $\ell_1$ and $\bar\ell_2$)

 \begin{itemize}
 \item[i)]   $pp\to (W^-\to \ell_1 \bar{\nu}_{\ell_1})(W^+\to \bar\ell_2 \nu_{\ell_2} )(VV\to 4j)$
 has a small cross-section,  about 0.04 fb~\cite{MADGRAPH} after imposing the cuts of eq.~(\ref{gencuts}),
 and can be reduced down to a negligible level by requring a below-treshold $\slashed{E}_T$. 

\item[ii)] $pp\to (W^-\to \ell_1 \bar{\nu}_{\ell_1})(W^+\to \bar\ell_2 \nu_{\ell_2} ) 4j$
is analogous to the previous process, but with the jets produced from QCD rather than electroweak
processes. The cross section is $50$ fb~\cite{Piccinini} (only the cuts on jets are imposed), 
below the background iii), and can be reduced
in a way similar to the one we now discuss.
%Piccinini dice sigma (pp-> W+ W- 4j) = 4.5 +- 0.2 pb.    Il BR e'  0.32^2/9   (1/9:  per singolo flavor)

\item[iii)] $pp\to  (\bar t\to \bar b  \ell_1 \bar{\nu}_{\ell_1}) (t   \to b \bar\ell_2 \nu_{\ell_2} )2j$
with the jets produced by QCD. After imposing the cuts of eq.~(\ref{gencuts}) the   cross section is 7200 fb  \cite{MADGRAPH}:
we need to suppress it with appropriate cuts, taking into account the peculiarities of the signal.
First, the requirment that the $2jb\bar{b}$ system reconstructs a $VV$ pair 
(imposing the criterion of eq.\eq{jjW}) lessen this background to 250 fb.
Second, we can require that leptons have a large enough $p_T$, eq.~(\ref{hardlepton}).

Let us consider for example the case $M=250 \GeV$:
imposing $p_T^\ell>70\, \textrm{GeV}$  reduces the background to 36 fb,
to be compared with a signal cross section of 
$$240 \,{\rm fb} \times{{\rm BR}(N^+N^0\to ZW^+\ell_1\bar{\ell}_2)} \circa{<} 8\,{\rm fb}.$$
We have not optimized the cuts in $p_T^\ell$ and $p_T^j$, and not used cuts on $\slashed{E}_T$.
For  larger values of $M$, 
one has to wait for a larger integrated luminosity for a discovery
which would also imply a better understanding of the detector, allowing to reliably impose a cut on
the absence of missing transverse energy $\slashed{E}_T$.
% This integrated luminosity will be collected in a very short time after the LHC startup and the fact that this BG can be suppressed even not relying on the absence of missing transverse energy let us hope that this discovery could be in the reach of the poorly understood detectors available at the beginning of LHC data taking.
%In case of larger values of $M$ one has to wait for a larger integrated luminosity which would also imply a better understanding of the detector. 
A larger $M$ implies a more central production of $N$: for example
we assume $M=750$ GeV, we impose $\eta_{\ell,j}<2.5$,
and fix eq.~(\ref{hardlepton}) to $p_T^\ell>140 \textrm{GeV}$ yielding a background cross section of about 1.7 fb. 
Imposing  $ \slashed{E}_T < 50 \textrm{ GeV}$ (that seems a reasonable estimate for the resolution of a well understood detector)
the background gets reduced to about $0.25$ fb.
%140
%at M=750 GeV sigmaBR=0.87fb sigmaBRcut=0.30 fb versus a 2jttbar 1.69(8)fb; 
%210 versus a 2jttbar 0.27(4) fb
%Assuming that the detector will be understood well enough, one can impose that the event has a missing transverse energy not larger than the energy resolution of the detector.  For instance imposing  $ \slashed{E}_T < 50 \textrm{ GeV}$ the BG process has  a cross section equal to 0.25(3) fb. 
Under the same cuts and the same BR as above the maximal signal  cross section is $0.08$ fb.
%allowing  a 5$\sigma$ discovery with about 300/fb of integrated luminosity.
%[O SOLO  $p_{T,j}>20\textrm{ GeV},\, \eta_j<5$]
%the cross section is about 5000 fb~\cite{MADGRAPH},
%and can be reduced by about one order of magnitude
%by requiring that the invariant mass of the $2j$ 
%[OR 4j??] matches $M_V$ (as in the signal)
%and by another order of magnitude by requiring that the $\slashed{E}_T$ carried away by neutrinos
%is below the threshold of LHC detectors. 
%The QCD originated processes (LFVa-c) are more difficult to deal with and the exact value of $\delta$ is needed to establish how much they can be reduced. STO PENSANDO COME SI PUñ STIMARE UN VALORE DI DELTA REALISITCO. PER ESEMPIO STUDIANDO LA MISSING $E_T$ CHE SI ORIGINA DALLA IMPERFETTA RICOTRUZIONE DEI JET. INFATTI L'ENERGIA DEI JET é MISURATA CON UN ERRORE $\sigma(E)/E=0.03+0.5/\sqrt{E/\textrm{GeV}}$.

\item[iv)] $pp\to (\bar t\to \bar b  \ell_1 \bar{\nu}_{\ell_1}) (t   \to b \bar\ell_2 \nu_{\ell_2} )(V\to 2j) $ has a total cross section of about 13 fb \cite{MADGRAPH}. The process iii) has a larger rate even after the requirements on the four jets. As such iii) is always dominant and this process can safely be neglected.

%This process differs from iii) only for the presence of a weak vector as a source of a pair of jets. Therefore this background is expected to have the same (or even higher) degree of reducibility of iii) after the requirement that 4j reconstruct a ZW system. At that stage iii) has a larger rate than this process and we already proved it to be negligible, thus we can conclude that the process iv) is negligible as well.

\end{itemize}

%
%Unfortunately the (LFVd) process matrix element is not currently available in any code known by the authors. This means that an events sample cannot be produced and its cross section cannot be easily computed. However its cross section can be estimated and upperbounded by the cross section of $ pp \to (W^+\to l^+\nu_l )(W^-\to l^-\bar{\nu}_l)3j$, which amounts to about 80fb \cite{AlpGen}. This cross section is higher than the typical signal cross section however it can be reduced imposing that the four QCD jets recostruct a ZV pair. The same kind of requirement has been used to reduce (LFVc), which had a much higher cross section. For this reason we consider (LFVd) as negligible compared to (LFVc).

%Reconstruction of a ZW pair into the 4 jets and the small missing $E_T$ condition lessen this cross section by a factor that can be estimated as follows:
%\begin{itemize}
%\item the probability of 4j faking a WZ pair can be estimated from a sample of W4j events on which we require that the 4 jets can be divided in two pairs one matching the W mass and the other matching the Z mass within a $n\cdot\Gamma_{Z,W}$.
%\item the probability that the produced neutrinos carry a $mET<\delta$ can be estimated from a sample of $pp\to WW(Z\to 2j)2j$ produced with ALPGEN \cite{AlpGen}. The threshold $\delta$ has to be determined on the base of instrumental sources of missing $E_T$ like jet energy scale and other measurment errors. 
%\end{itemize}

%%%%%%%%%%%%%%%%%%%%%%%%%%%%%%%%%%%%%%%%%%%%%%%%%  
\subsection{Lepton-number violating signatures}\label{LNV}
Lepton-number-violation can be discovered in the channels
\beq \begin{array}{l}
pp\to (N^+\to  \ell^+_1 Z)(N^0\to \ell^+_2 W^-)= \ell^+_1 \ell^+_2 ZW^-\\
pp\to  (N^-\to \ell_1^-Z)(N^0\to \ell_2^-  W^+ )
= \ell_1^- \ell_2^- Z W^+\end{array}
 \label{eq:LV2}.\eeq
All the $N\to{\rm SM}$ decays in eq.s~(\ref{sys:N0decay}) and~(\ref{sys:Npmdecay})
are expected to have comparable sizable branching ratios
and measuring the flavors of $\ell_i$  would allow to identify the flavor of the lepton doublet
coupled to the triplet $N$.
%{\bf If $\tilde{m} \circa{<} 10^{-3}\,\hbox{eV}\cdot (100 \,\hbox{GeV}/M)^2$ the $(N^\pm \to N^0 \pi^\pm )$   is important. In this case the most interesting channels are the lepton number violating ones:}
If $N^\pm \to N^0 \pi^\pm $ also happen to have a detectably large branching,
% {\bf ratio, which occurs, as explained above, if $\tilde{m}$ is small enough}, 
extra $L$-violating channels
where a soft $\pi^\pm$ is present arise:
\beq pp \to  \ell_1\ell_2 \pi^+ Z W^+  ,\qquad \ell_1\ell_2 \pi^\pm W^+ W^+ ,\qquad
\ell_1\ell_2 \pi^+\pi^- W^+ W^+.\label{eq:pichannel}\eeq
Even if the $\pi^\pm$ are too soft to be detected, 
BR$(N^\pm \to N^0 \pi^\pm )$ can still be measured from the relative
rate of $\ell_1 \ell_2 W^+ W^+$ vs $\ell_1 \ell_2 ZW^+$ events
(using decay channels that allow to discriminate a  $Z$ from a $W^\pm$; 
$Z \to b \bar b$ or into leptons allow an  event-by-event discrimination)
%{\bf In this case the measurement of the BR of decays with and without pion, possible if one BR is not much larger than the other one) would allow to infer the value of $\tilde{m}_1$.}
allowing to infer the value of $\tilde{m}_1$: both type of decays have a detectably large BR 
if  $\tilde{m}_1\sim 10^{-(3-4)}\,\hbox{eV}\cdot (100 \,\hbox{GeV}/M)^2$.
Neglecting the soft $\pi^\pm$, all the final-state particles of eq.s\eq{LV2} and\eq{pichannel}
can be seen, allowing for a precise measurement of $M$.

%Channels involving missing transverse energy in neutrinos, such as
%$\ell \bar\nu ZZ$  and $\ell\nu ZZ$ can also indirectly show the existence of the $L$-violating channel.

Let us now discuss the backgrounds 
\begin{itemize}
\item[i)] The physical background to\eq{LV2} is 
$pp\to VV(W^\pm\to   \ell^\pm_1 \nubarnu)(W^\pm\to  \ell^\pm_1 \nubarnu)$ where $V$ are heavy SM vectors and the (anti-)neutrinos  happen to carry so little
missing transverse energy that their presence is not detected.
Using {\sc MadGraph}~\cite{MADGRAPH} we find, at LHC
$\sigma(pp\to VVW^+W^+)\approx 0.8\textrm{ fb}$ and $\sigma(pp\to VVW^-W^-)\approx 1\textrm{ fb}$.
%$\sigma(pp\to W^+ W^+ W^- W^-) \approx 0.62\,{\rm fb}$,
%$\sigma(pp\to W^+ W^- W^- Z) \approx 0.22\,{\rm fb}$ and
%$\sigma(pp\to W^+ W^- Z Z)\approx 0.48 \,{\rm fb}$.
%For example, the signal $pp\to \ell_1 \ell_2 W^+ W^+$ can be faked by $W^+ W^+ W^- W^-$ with the two $W^-$ decaying into $\ell_1$ and $\ell_2$ producing two anti-neutrinos that happen to carry so little missing transverse energy that their presence is not detected.
%So one expects a background cross section well below the $0.1\,{\rm fb}$ level,
%that can be further suppressed by requiring hard leptons. 
%searching peaks at invariant mass $M$.
\end{itemize}
As in the previous section, 
hadronic decays of the $W^\pm, Z$ give the signal with the higher rate:
\beq
pp \to \ell_1 \ell_2 4j \label{eq:LV}
\eeq
without missing transverse energy $E_T$.
The cross section of the already discussed processes i)
$pp\to (W^-\to \ell_1 \bar{\nu}_{\ell_1})(W^-\to\ell_2  \bar\nu_{\ell_2} )(VV\to 4j)$
is about 0.05 fb, which is already negligible before imposing the cuts of eq.~(\ref{gencuts}), (\ref{hardlepton}), etc.
The other background to\eq{LV} is
\begin{itemize}
\item[ii)] 
$pp\to (W^-\to \ell_1 \bar{\nu}_{\ell_1})(W^-\to\ell_2  \bar \nu_{\ell_2} ) 4j$
where the jets are produced from QCD.
The cross section is $20$ fb~\cite{Piccinini} (summed over lepton flavors; only cuts on jets are imposed),
% Piccinini dice 200 fb per pp -> W+W+4j e che pp-> W-W-4j ha sigma molto simile.  Poi ci sono i BR dei W.
and can be significantly reduced in a way similar to background iii) in section~\ref{sec:LFV}.
\end{itemize}
Notice that the lepton-number-violating signature also allows to see lepton-flavour violation
and, compared to the lepton-flavour violating and lepton-number conserving signal of section~\ref{sec:LFV},
has a comparable rate and a much less background, because it cannot be faked by $t\bar t$ production.
An integrated luminosity of 10/fb should allow to see the LNV signal for $M \circa{<}0.8\TeV$.
While the LFV signal of  section~\ref{sec:LFV} should not be used for a discovery search, its later detection would
clarify the physics: e.g.\ it is produced by type-III see-saw and not by type-II see-saw.

\medskip

\begin{figure}[t]
\begin{center}
\includegraphics[width=\textwidth]{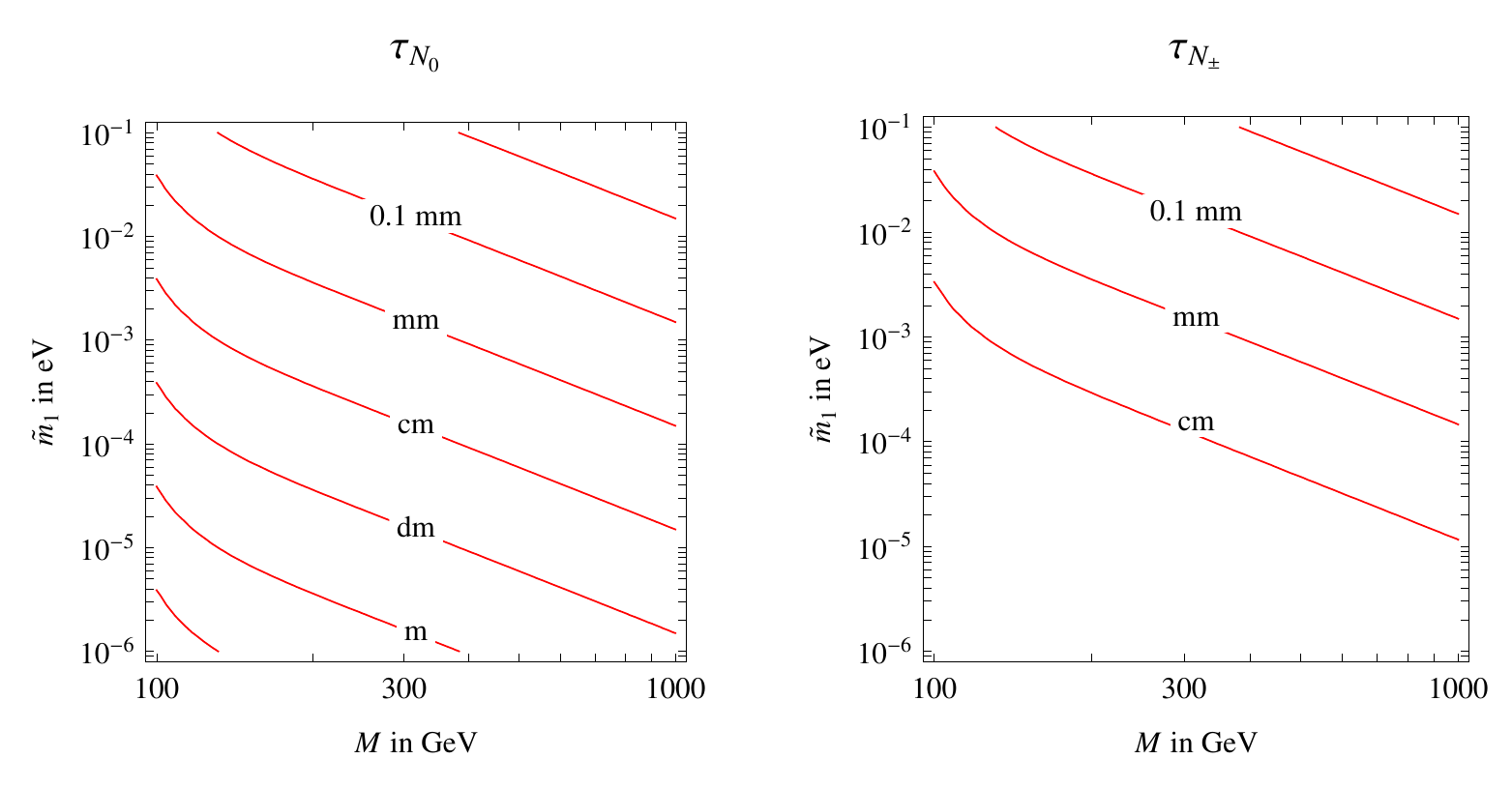}
\caption{\label{fig:tau}\em Contour-plot of triplet  life-times for $m_h=115\GeV$.}
\end{center}
\end{figure}

\begin{figure}[t]
\begin{center}
\includegraphics[width=\textwidth]{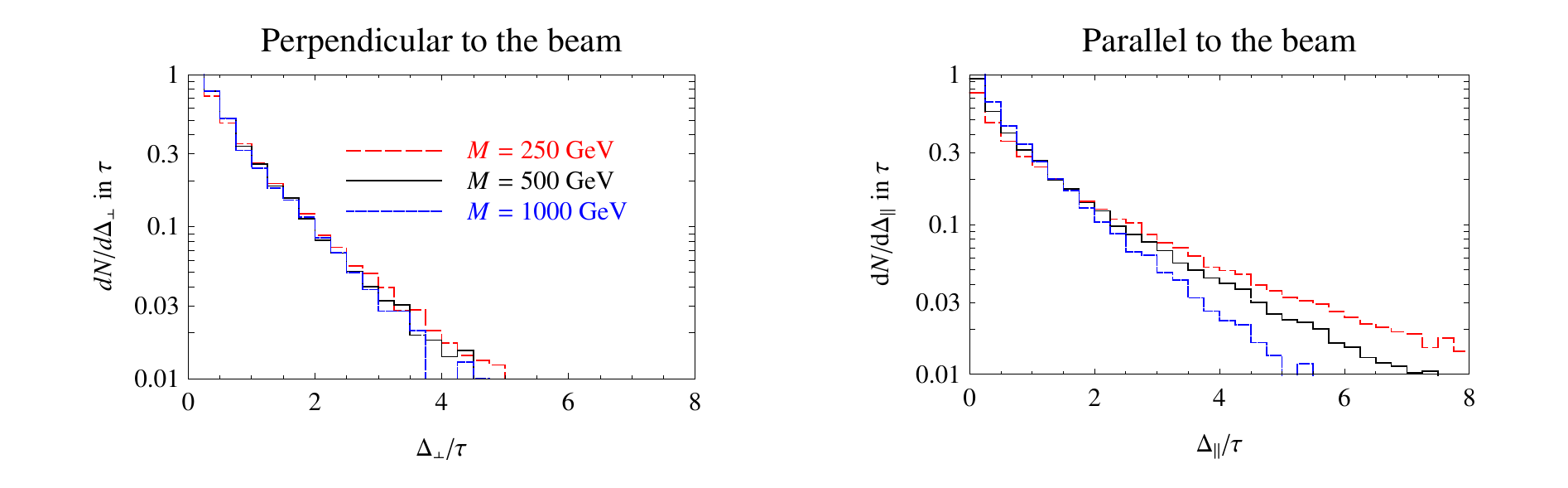}
\caption{\label{fig:vertice}\em Distribution of the displacement of the secondary vertex from the interaction point.}
\end{center}
\end{figure}

%%%%%%%%%%%%%%%%%%%%%%%%%%%%%%%%%%%%%%%%%%%%%%%
\subsection{Displaced vertices}
Moreover all the channels above can have an extra signature:
we observe that the lifetimes
$\tau_{N^a} = 1/\sum_f \Gamma(N^a\to f)$ can be so large
that $N_0,N_\pm$-decay vertices can be detectably displaced from the primary production vertex.  
Indeed in the SU(2)-invariant limit $M\gg M_Z$ one has
\beq \tau_{N_0} = \tau_{N_\pm} = \frac{8\pi v^2}{\tilde{m}_1 M^2}=1.5\cm
\frac{\rm meV}{\tilde{m}_1} \Big(\frac{100\GeV}{M}\Big)^2\label{eq:displacedvertex}.\eeq
%and fig.\fig{Gamma} shows the exact dependence of all decay widths $\Gamma$ on $M$.
Fig.\fig{tau} shows the total life-times  $\tau$ of $N^0$ and of $N^\pm$ as function of $M$ and $\tilde{m}_1$.
We see that there are two possible cases.
\begin{enumerate}
\item For larger $\tilde{m}_1$, one has smaller $\tau_{N^0}\approx \tau_{N^\pm}$: e.g.\
$\tau_{N^0}\approx \tau_{N^\pm} \approx 0.3 \,{\rm mm}\cdot (M/100\,\hbox{GeV})^2$ for
$\tilde{m}_1 \approx (\Delta m^2_{\rm atm})^{1/2}\simeq 0.05$ eV. 

\item For smaller $\tilde{m}_1$ one has $\tau_{N^0}\gg  \tau_{N^\pm}\sim 5$ cm;
$N^\pm$ decays predominantly to $N^0 \pi^\pm$ leading to multiple displaced vertices. 
Unfortunately the $\pi^\pm$ produced in the $N^\pm$  decay are too soft to be detected and the typical track produced by $N^\pm$ seems too short to be % reconstructed and 
well measured. 
%It can be tagged by the $N^0$ decay, if it takes place close enough.

\end{enumerate}
Fig.s\fig{vertice}a and b show the distribution in the secondary vertex displacement $\Delta$ for triplets produced at LHC,
after taking into account the time-dilatation effect.
We see that the average displacement perpendicular to the beam axis is
$\langle \Delta_\perp\rangle \approx 0.9\tau$,
with a minor dependence on $M$.
In the direction parallel to the beam axis one has $\langle \Delta_\parallel\rangle \approx 2.4\tau$
at $M\approx 250\GeV$ and  $\langle \Delta_\parallel\rangle \approx \tau$ at $M=1\TeV$.
Both distributions are very roughly exponentials,  $dN/d\Delta \approx e^{-\Delta/\langle \Delta\rangle}$.

Capabilities of LHC detectors (ATLAS, CMS%
%and LHCb
 )
strongly depend on the unknown
flavor composition of the lepton coupled to $N$ and
on the displacement $\Delta$, because
decays would happen in different parts of the apparatus.
For smaller $\Delta$, 
LHC detectors should allow to reconstruct the position of the secondary vertex
with an uncertainty of about $0.5\,{\rm mm}$ and $0.1\,{\rm mm}$, in the directions 
parallel and orthogonal to the beam axis respectively~\cite{LHCsecondary}. 
For larger $\Delta$, the $N_0$ displacement can be $\Delta_\perp\circa{>}50\cm$:  in such a case LHC detectors could 
see the muons but not electrons and taus which need almost the full lenght of the inner detector to be properly observed; although no dedicated studies exist,
a displaced vertex could be identified if $N_0\to \mu^\pm W^\mp$ happens
around the first layers of the muon detector, and the $W$ decays hadronically.
%The decays  $N^\pm\to \mu^\pm Z \to \mu^\pm \mu^-\mu^+$ should allow
%to reconstruct the transverse displacement with an uncertainty possibly better than $0.1\,{\rm mm}$
%if these decays occurs sufficently inside the tracker (typically for $\tau \circa{<}  1$ m) or with an uncertainty of few 0.1 mm from the $\mu$ calorimeter  for $\tau \circa{<} (5-10)$ m.
%%unless the displacement is bigger than a few meters.
%A few times worse performance should be achieved for channels involving electrons observed in the tracker. In principle electrons created far from the collision point could be observed in the muon calorimeter but with rather poor angular resolution. 
%The situation is more difficult for taus, although one can rely on tau decays into muons.

Finally, if $\Delta \circa{>} 10\,{\rm m}$, all the production channels of eq. (\ref{eq:V}) result in the effective production of $N^0N^0$ plus one or two undetectable pions. The decayed $N^\pm$ have too a short track to be tagged or well measured, thus they are of no help. The produced $N^0$s escape the detector and therefore the event has no trigger and no signature. For this reason the detection of events with such a large $\Delta$ seems very challenging. 

%Finally, if $\Delta \circa{>} 10\,{\rm m}$, the two $N^0$ carry away some missing transverse energy, 
%without leaving almost any signature that allows to trigger the event.

\medskip

This opens a plethora of scenarios.
In view of the first  inequality in eq.\eq{boundsum} (that holds with three triplets), any observed triplet lifetime larger than 
$0.3\, \hbox{mm} \cdot (100 \GeV/M)^2$ would imply a small $m_1 <(\Delta m^2_{\rm atm})^{1/2}$ i.e.\
non-degenerate neutrino spectrum.
In the most optimistic cases where all triplets ($N\equiv N_1$
and $N_2,N_3$)
have sub-TeV masses, one could infer informations on the neutrino masses and mixings
and test the consistency of the scenario via the bounds of eq.\eq{boundsum}.
E.g.\ the second inequality in eq.\eq{boundsum} implies
%For example, eq.~(\ref{eq:boundsum}) implies 
max$(\tilde{m}_1,\tilde{m}_2,\tilde{m}_3) \geq  \sum_i m_{i}/3$ or equivalently 
\beq \tau_{\rm min}\equiv{\rm min}\,(\tau_1,\tau_2,\tau_3) \leq 1\, \hbox{mm} \cdot (0.05\,\hbox{eV}/\sum_i m_{i}) (100 \,\hbox{GeV}/M)^2.\label{eq:displacedhier}\eeq
Measuring $\tau_{\rm min}\circa{<} 1\,{\rm mm}$ and $M \approx 100\GeV$
would e.g.\ point to neutrinos with normal mass hierarchy
($\sum m_i \approx 0.05\eV$) since inverted mass hierarchy ($\sum m_i \approx 0.1\eV$) or
a quasi-degenerate spectrum ($\sum m_i > 0.15\eV$) lead to a stronger upper limit on $\tau_{\rm min}$.
Therefore the type-III see-saw allows to measure the couplings directly related to neutrino mass physics from displaced vertices. 
\medskip

\subsection{Comparison with displaced vertices in other seesaw models}
In the type-I see-saw, right-handed neutrinos are negligibly produced unless some
Yukawa coupling is much larger than what suggested by neutrino masses
(thereby loosing a direct connection between LHC and neutrino physics)
but such a large Yukawa would not lead to displaced vertices.
Alternatively, right-handed neutrinos might be produced by some extra interaction
(such as gauged U(1)$_{B-L}$  
%and/or $\SU(2)_L$ extended to $\SU(2)_L\otimes\SU(2)_R$ 
and/or $\SU(2)_L$ extended to $\SU(3)_L$)\footnote{Production from $\SU(2)_L$ extended to
$\SU(2)_L\otimes \SU(2)_R$ would not lead to displaced $N$ vertices due to fast $N$ decays mediated
by a right-handed gauge boson $W_R^\pm$, such as $N \rightarrow \ell_R u_R \bar{d}_R$.}: it must be stressed that the $pp \rightarrow N^0 N^0 \rightarrow \ell_1^\pm \ell_2^\pm W^\mp W^\mp$ or $pp \rightarrow N^0 N^0 \rightarrow \ell_1^\pm \ell_2^\mp W^\mp W^\pm$ channels would lead to a displaced vertex
phenomenology similar to the one above, eq.s\eq{displacedvertex} and\eq{displacedhier}. The
lepton-number violating $N^0N^0 \rightarrow \ell_1 \ell_2 W^+ W^+$ channel is of particular interest.
This signal is also generated by the scalar triplet $T$ of type II see-saw,
together with\eq{LV}, and together with lepton-number conserving $\ell_1 \ell_2 \bar \ell_3\bar\ell_4$
and with $W^+ W^+ W^-W^-$~\cite{LHCtypeII}.
%In the triplet model above, production of $N^0N^0$ is also possible but is loop suppressed and was neglected.
%$N^0N^0$ decays lead to the lepton-number violating $\ell_1 \ell_2 W^+ W^+$ signal.

Type-II see-saw can also lead to displaced vertices, but of limited size: the life-time $\tau_T$ of the scalar triplet $T$ depends on its unknown relative branching ratios into lepton or Higgs pairs, and is maximal when the two BR are equal~\cite{AHHRR}: 
$\tau_T < 8 \pi v^2 / (M^2_T \sqrt{\sum_i m_i^2})$. This leads to  $\tau_T< 0.3$ mm $(M_T/100\,\hbox{GeV})^2$ for normal light neutrino hierarchy, $\tau_T< 0.15$ mm $(M_T/100\,\hbox{GeV})^2$ for inverted hierarchy
and less for degenerate spectrum. Consequently, here too the hierarchy could in principle be distinguished, but only for approximately equal two branching ratios.

\section{Conclusions}
Although it seems unlikely  that neutrino masses are generated by the type-III see-saw at energies accessible to LHC,
we studied what such an encounter of the third type would look like,
finding very characteristic signals that allow to reconstruct the Lagrangian.

The unknown parameters are: the mass $M$ of the lightest fermion triplet $N=\{N^0,N^\pm\}$
(other triplets might  be accessible at LHC);
its  contribution $\tilde{m}_1$ to neutrino masses,
and the lepton flavor coupled to $N$.
Triplet gauge couplings are fully predicted,
and they induce a small mass splitting
$M_{N^\pm} - M_{N^0}\approx 166\MeV$ among the charged $N^\pm$ and neutral $N^0$ components of the triplet $N$.
Production of triplets in $pp$ collisions is dominated by gauge couplings,
and fig.\fig{sigma} shows the predicted cross sections for $pp\to N^+ N^-, N^+ N^0,N^- N^0$
($N^0 N^0$ pairs are not produced)
as function of the triplet mass $M$.
This mass can be inferred from the cross section,
from the  distribution in transverse momentum, 
and measured from the invariant mass of appropriate final-state particles.
At an $e^- e^+$ collider, the cross section is maximal at $s =1.4\times 4M^2$,
and for the same reason triplets produced by $q\bar{q}'$ partonic collisions are typically
moderately relativistic.

Pair production of triplets and their decays gives lepton-flavor and lepton-number violating signals.
There are two kinds of decays: 
\begin{itemize}
\item[i)] $N^0,N^\pm$ into a lepton plus a massive SM vector (or higgs, if it is lighter than the triplet),
with width $\Gamma \sim \tilde{m}_1 M^2/v^2$, that is small enough to lead to detectably displaced
decay vertices if $\tilde{m}_1 \circa{<} 0.1\eV (M/100\, \hbox{GeV})^2$.
\item[ii)] $N^\pm \to N^0 \pi^\pm$, with width $\Gamma \sim 1/(\hbox{few cm})$.
\end{itemize}
In section~\ref{highest} we studied the signal with the higher rate,
$pp\to 4j\ell^\pm\slashed{E}_T$;
in sections~\ref{sec:LFV} and~\ref{LNV} we studied the most characteristic final states
that violate lepton flavor and lepton number:
two charged leptons accompanied by two massive vectors or higgs.
In all cases we studied the SM backgrounds and how they can be suppressed allowing for a discovery.
The most promising discovery channel seems $pp\to \ell_1^+ \ell_2^+4j$, as it has a small background and
its rate is only a factor 2 below the higher rate.

The decays i) can allow to measure the lepton flavor coupled to $N$.
%give the lepton-number-violating signal $pp\to \ell_1\ell_2 Z W^+$ and $\ell^+_1 \ell^+_2 Z W^-$.
%Together with the decays ii), one also get the signal $\ell_1 \ell_2 W^+ W^+$ and $\ell^-_1 \ell^-_2 Z W^-$
%accompanied by one or two pions with energy $\sim 100\MeV$.
Finally, the parameter $\tilde{m}_1$ can be inferred from either the BR of the decay mode ii),
or from measuring how much secondary decay vertices are displaced from the production point.
These two signals allow to roughly  cover the range
$10^{-7}\eV\circa{<}\tilde{m}_1 \circa{<}0.1\eV$, especially if $M$ is as light as possible.

\paragraph{Acknowledgements} 
A.S.\ thanks Milla Baldo Ceolin for having prompted this paper. T.H. thanks the FRS-FNRS for support.
We thank Fulvio Piccinini for help with $2V4j$ backgrounds.


\small

\begin{thebibliography}{nn}
\bibitem{review}
For recent reviews see:
\hepart[hep-ph/0606054]{A.~Strumia and F.~Vissani}.
\hepart[0704.1800]{M.C. Gonzalez-Garcia, M. Maltoni}.



\bibitem{typeI}
T.~Han and B.~Zhang,
  %``Signatures for Majorana neutrinos at hadron colliders,''
  Phys.\ Rev.\ Lett.\  {97} (2006) 171804
  [arXiv:hep-ph/0604064].
 F. del Aguila, J.A. Aguilar-Saavedra and R. Pittau, JHEP 0710 (2007) 047 [hep-ph/0703261].
J.~Kersten and A.~Y.~Smirnov,
  %``Right-Handed Neutrinos at LHC and the Mechanism of Neutrino Mass
  %Generation,''
  Phys.\ Rev.\  D {76} (2007) 073005
  [arXiv:0705.3221 [hep-ph]].


\bibitem{LHCtypeIIa}
M. M\"uhlleitner and Michael Spira, Phys. Rev. {D68} (2003) 117701 [arXiv:hep-ph/0305288];
   A.~G.~Akeroyd and M.~Aoki,
 %``Single and pair production of doubly charged Higgs bosons at hadron
 %colliders,''
 Phys.\ Rev.\  D {72} (2005) 035011
 [arXiv:hep-ph/0506176];
 E.~Accomando {\it et al.},
 %``Workshop on CP studies and non-standard Higgs physics,''
 arXiv:hep-ph/0608079;
 K.~Huitu, J.~Maalampi, A.~Pietila and M.~Raidal,  %``Doubly charged Higgs at LHC,''
  Nucl.\ Phys.\  B {487} (1997) 27; A.~Hektor, M.~Kadastik, M.~Muntel, M.~Raidal and L.~Rebane, %``Testing neutrino masses in little Higgs models via discovery of doubly
 %charged Higgs at LHC,''
  arXiv:0705.1495 [hep-ph];
E.~J.~Chun, K.~Y.~Lee and S.~C.~Park,
 %``Testing Higgs triplet model and neutrino mass patterns,''
 Phys.\ Lett.\  B {566} (2003) 142
 [arXiv:hep-ph/0304069].


\bibitem{LHCtypeII}
\hepart[0803.3450]{P.~Fileviez Perez, T.~Han, G.~Y.~Huang, T.~Li and K.~Wang}.
\hepart[0804.1265]{W.~Chao, Z.~G.~Si, Z.~z.~Xing and S.~Zhou}

\bibitem{foot}
R. Foot, H. Lew, X.-G. He and G.C. Joshi,
  Z. Phys. {C44} (1989) 441; 


\bibitem{ma1} E.~Ma,
 %``Pathways to naturally small neutrino masses,''
 Phys.\ Rev.\ Lett.\  {81} (1998) 1171
 [arXiv:hep-ph/9805219]; 

\bibitem{bajc1}  B.~Bajc and G.~Senjanovic,
 %``Seesaw at LHC,''
 JHEP {0708} (2007) 014
 [arXiv:hep-ph/0612029]; I.~Dorsner and P.~Fileviez Perez,
%``Upper Bound on the Mass of the Type III Seesaw Triplet in an SU(5) Model,''
 JHEP {0706} (2007) 029
 [arXiv:hep-ph/0612216].
 %%CITATION = JHEPA,0708,014;%%


  \bibitem{ABBGH2} 
\hepart[0803.0481]{A. Abada et al.}
  
\bibitem{BP1} W. Buchm\"uller and M. Plumacher, Phys. Rept {320} (1999) 329.
%The bound $\tilde{m}_1\ge m_1$ holds up to RGE corrections, that
%can generate a 


 \bibitem{LeptogenesisTypeIII}
\art[hep-ph/0312203]{T. Hambye et al.}{Nucl. Phys.}{B695}{169}{2004}.


\bibitem{shap}
Y.~Burnier, M.~Laine and M.~Shaposhnikov,
 %``Baryon and lepton number violation rates across the electroweak
 %crossover,''
 JCAP {0602} (2006) 007
 [arXiv:hep-ph/0511246].
 
 

\bibitem{Hall} 
See e.g.
\hepart[0712.2454]{L.~J.~Hall and Y.~Nomura}.

 
\bibitem{MDM}
 \art[hep-ph/0512090]{M. Cirelli, N. Fornengo, A. Strumia}{Nucl. Phys.}{B753}{178}{2006}.


 \bibitem{Bajc} B.~Bajc, M.~Nemevsek and G.~Senjanovic,
 %``Probing seesaw at LHC,''
 Phys.\ Rev.\  D {76} (2007) 055011
 [arXiv:hep-ph/0703080].

\bibitem{STWY}
\art[hep-ph/0405040]{R.~Barbieri, A.~Pomarol, R.~Rattazzi and A.~Strumia}{Nucl.\ Phys.\ B}{703}{2004}{127}.
The correction to $W$ is the same that applies in supersymmetric models in the limit of pure wino
(the wino has the same quantum numbers as the triplet of type III see-saw,
but supersymmetry forbids the small Yukawa coupling in eq.\eq{Lseesaw3}):
\art[hep-ph/0502095]{G.~Marandella, C.~Schappacher and A.~Strumia}{Nucl.\ Phys.\  B}{715}{2005}{173}.


\bibitem{Ma}  E.~Ma and D.~P.~Roy,
 %``Heavy triplet leptons and new gauge boson,''
 Nucl.\ Phys.\  B {644} (2002) 290
 [arXiv:hep-ph/0206150].


 \bibitem{pdf} A.~D.~Martin, R.~G.~Roberts, W.~J.~Stirling and R.~S.~Thorne,
 %``Uncertainties of predictions from parton distributions. I: Experimental
 %errors. ((T)),''
 Eur.\ Phys.\ J.\  C {28} (2003) 455
 [arXiv:hep-ph/0211080].
We used the MRST 2002 parton distributions
 as extracted from \url{http://durpdg.dur.ac.uk/hepdata/pdf3.html}
 and converted into Mathematica form. 
We also used  
 S.~Alekhin, K.~Melnikov and F.~Petriello,
 %``Fixed target Drell-Yan data and NNLO QCD fits of parton distribution
 %functions,''
 Phys.\ Rev.\  D {74} (2006) 054033
 [arXiv:hep-ph/0606237],
 available in  Mathematica form thanks to Sergey Kulagin,
 web site \url{http://sirius.ihep.su/~alekhin/pdfa02}.


\bibitem{MADGRAPH}
{\sc MadGraph},
F.~Maltoni and T.~Stelzer,
%``MadEvent: Automatic event generation with MadGraph,''
JHEP {0302}, 027 (2003)
[hep-ph/0208156].
%%CITATION = JHEPA,0302,027;%%
%\cite{Stelzer:1994ta}
%\bibitem{Stelzer:1994ta}
T.~Stelzer and W.~F.~Long,
%``Automatic generation of tree level helicity amplitudes,''
Comput.\ Phys.\ Commun.\  {81}, 357 (1994)
[hep-ph/9401258].
%%CITATION = CPHCB,81,357;%%
%\cite{Alwall:2007st}
%\bibitem{Alwall:2007st}
J.~Alwall {\it et al.},
%``MadGraph/MadEvent v4: The New Web Generation,''
JHEP {0709}, 028 (2007)
[0706.2334 [hep-ph]].
%Web page: \myurl{madgraph.phys.ucl.ac.be}{madgraph.phys.ucl.ac.be}.

\bibitem{AlpGen}
{\sc AlpGen},
\art[hep-ph/0206293]{M.L. Mangano, M. Moretti, F. Piccinini, R. Pittau, A. Polosa}{JHEP}{0307}{001}{2003}.

\bibitem{Piccinini}
F. Piccinini, private communication.
This  process will be implemented in a future release of {\sc AlpGen}~\cite{AlpGen},
allowing to precisely design cuts.




\bibitem{LHCsecondary} 
%LHCsecondary [0803.4405].
We thank Pascal Vanlaer,  Andrea Perrotta, Roberto Tenchini for discussions about capabilities of LHC detectors.


\bibitem{AHHRR} G. D'Ambrosio et al., 
Phys. Lett. {B604} (2004) 199
[arXiv:hep-ph/0407312]. 

\end{thebibliography}
\end{document}